\documentclass[aps,prd,twocolumn,superscriptaddress,nofootinbib]{revtex4-2}

\usepackage{amsmath}
\usepackage{amssymb}
\usepackage{natbib}
\usepackage{graphicx}
\usepackage{color}
\usepackage{hyperref}
\usepackage{lineno}
\usepackage{booktabs}
\usepackage{multirow}
\usepackage{orcidlink}

\usepackage{savesym}
\savesymbol{tablenum}
\usepackage{siunitx}
\restoresymbol{SIX}{tablenum}

\DeclareMathOperator{\sinc}{sinc}

\begin{document}

\newcommand{\fixme}[1]{\textcolor{red}{FIXME: #1}}
\newcommand{\refresponse}[1]{\textcolor{red}{#1}}
\let\svthefootnote\thefootnote
\newcommand\blankfootnote[1]{%
  \let\thefootnote\relax\footnote{#1}%
  \addtocounter{footnote}{-1}\let\thefootnote\svthefootnote
}
\title{Searching for axionlike time-dependent cosmic birefringence with data from SPT-3G}

\affiliation{Department of Physics and Astronomy, University of California, Los Angeles, CA, 90095, USA}
\affiliation{Fermi National Accelerator Laboratory, MS209, P.O. Box 500, Batavia, IL, 60510, USA}
\affiliation{Kavli Institute for Cosmological Physics, University of Chicago, 5640 South Ellis Avenue, Chicago, IL, 60637, USA}
\affiliation{Department of Astronomy and Astrophysics, University of Chicago, 5640 South Ellis Avenue, Chicago, IL, 60637, USA}
\affiliation{Department of Physics and Astronomy, Michigan State University, East Lansing, MI 48824, USA}
\affiliation{School of Physics and Astronomy, Cardiff University, Cardiff CF24 3YB, United Kingdom}
\affiliation{Department of Astronomy, University of Illinois Urbana-Champaign, 1002 West Green Street, Urbana, IL, 61801, USA}
\affiliation{Center for AstroPhysical Surveys, National Center for Supercomputing Applications, Urbana, IL, 61801, USA}
\affiliation{Department of Physics, University of California, Berkeley, CA, 94720, USA}
\affiliation{School of Physics, University of Melbourne, Parkville, VIC 3010, Australia}
\affiliation{Institut d'Astrophysique de Paris, UMR 7095, CNRS \& Sorbonne Universit\'{e}, 98 bis boulevard Arago, 75014 Paris, France}
\affiliation{High-Energy Physics Division, Argonne National Laboratory, 9700 South Cass Avenue., Lemont, IL, 60439, USA}
\affiliation{Kavli Institute for Particle Astrophysics and Cosmology, Stanford University, 452 Lomita Mall, Stanford, CA, 94305, USA}
\affiliation{Department of Physics, Stanford University, 382 Via Pueblo Mall, Stanford, CA, 94305, USA}
\affiliation{SLAC National Accelerator Laboratory, 2575 Sand Hill Road, Menlo Park, CA, 94025, USA}
\affiliation{Enrico Fermi Institute, University of Chicago, 5640 South Ellis Avenue, Chicago, IL, 60637, USA}
\affiliation{Department of Physics, University of Chicago, 5640 South Ellis Avenue, Chicago, IL, 60637, USA}
\affiliation{High Energy Accelerator Research Organization (KEK), Tsukuba, Ibaraki 305-0801, Japan}
\affiliation{Department of Physics and McGill Space Institute, McGill University, 3600 Rue University, Montreal, Quebec H3A 2T8, Canada}
\affiliation{Canadian Institute for Advanced Research, CIFAR Program in Gravity and the Extreme Universe, Toronto, ON, M5G 1Z8, Canada}
\affiliation{Joseph Henry Laboratories of Physics, Jadwin Hall, Princeton University, Princeton, NJ 08544, USA}
\affiliation{Department of Astrophysical and Planetary Sciences, University of Colorado, Boulder, CO, 80309, USA}
\affiliation{Department of Physics, University of Illinois Urbana-Champaign, 1110 West Green Street, Urbana, IL, 61801, USA}
\affiliation{Department of Physics, Case Western Reserve University, Cleveland, OH, 44106, USA}
\affiliation{CASA, Department of Astrophysical and Planetary Sciences, University of Colorado, Boulder, CO, 80309, USA }
\affiliation{Department of Physics, University of Colorado, Boulder, CO, 80309, USA}
\affiliation{Department of Physics \& Astronomy, University of California, One Shields Avenue, Davis, CA 95616, USA}
\affiliation{Physics Division, Lawrence Berkeley National Laboratory, Berkeley, CA, 94720, USA}
\affiliation{Materials Sciences Division, Argonne National Laboratory, 9700 South Cass Avenue, Lemont, IL, 60439, USA}
\affiliation{California Institute of Technology, 1200 East California Boulevard., Pasadena, CA, 91125, USA}
\affiliation{Three-Speed Logic, Inc., Victoria, B.C., V8S 3Z5, Canada}
\affiliation{David A. Dunlap Department of Astronomy \& Astrophysics, University of Toronto, 50 St. George Street, Toronto, ON, M5S 3H4, Canada}
\author{K.~R.~Ferguson*\,\orcidlink{0000-0002-4928-8813}}
\affiliation{Department of Physics and Astronomy, University of California, Los Angeles, CA, 90095, USA}
\author{A.~J.~Anderson\,\orcidlink{0000-0002-4435-4623}}
\affiliation{Fermi National Accelerator Laboratory, MS209, P.O. Box 500, Batavia, IL, 60510, USA}
\affiliation{Kavli Institute for Cosmological Physics, University of Chicago, 5640 South Ellis Avenue, Chicago, IL, 60637, USA}
\affiliation{Department of Astronomy and Astrophysics, University of Chicago, 5640 South Ellis Avenue, Chicago, IL, 60637, USA}
\author{N.~Whitehorn\,\orcidlink{0000-0002-3157-0407}}
\affiliation{Department of Physics and Astronomy, Michigan State University, East Lansing, MI 48824, USA}
\affiliation{Department of Physics and Astronomy, University of California, Los Angeles, CA, 90095, USA}
\author{P.~A.~R.~Ade}
\affiliation{School of Physics and Astronomy, Cardiff University, Cardiff CF24 3YB, United Kingdom}
\author{M.~Archipley\,\orcidlink{0000-0002-0517-9842}}
\affiliation{Department of Astronomy, University of Illinois Urbana-Champaign, 1002 West Green Street, Urbana, IL, 61801, USA}
\affiliation{Center for AstroPhysical Surveys, National Center for Supercomputing Applications, Urbana, IL, 61801, USA}
\author{J.~S.~Avva}
\affiliation{Department of Physics, University of California, Berkeley, CA, 94720, USA}
\author{L.~Balkenhol\,\orcidlink{0000-0001-6899-1873}}
\affiliation{School of Physics, University of Melbourne, Parkville, VIC 3010, Australia}
\author{K.~Benabed}
\affiliation{Institut d'Astrophysique de Paris, UMR 7095, CNRS \& Sorbonne Universit\'{e}, 98 bis boulevard Arago, 75014 Paris, France}
\author{A.~N.~Bender\,\orcidlink{0000-0001-5868-0748}}
\affiliation{High-Energy Physics Division, Argonne National Laboratory, 9700 South Cass Avenue., Lemont, IL, 60439, USA}
\affiliation{Kavli Institute for Cosmological Physics, University of Chicago, 5640 South Ellis Avenue, Chicago, IL, 60637, USA}
\author{B.~A.~Benson\,\orcidlink{0000-0002-5108-6823}}
\affiliation{Fermi National Accelerator Laboratory, MS209, P.O. Box 500, Batavia, IL, 60510, USA}
\affiliation{Kavli Institute for Cosmological Physics, University of Chicago, 5640 South Ellis Avenue, Chicago, IL, 60637, USA}
\affiliation{Department of Astronomy and Astrophysics, University of Chicago, 5640 South Ellis Avenue, Chicago, IL, 60637, USA}
\author{F.~Bianchini\,\orcidlink{0000-0003-4847-3483}}
\affiliation{Kavli Institute for Particle Astrophysics and Cosmology, Stanford University, 452 Lomita Mall, Stanford, CA, 94305, USA}
\affiliation{Department of Physics, Stanford University, 382 Via Pueblo Mall, Stanford, CA, 94305, USA}
\affiliation{SLAC National Accelerator Laboratory, 2575 Sand Hill Road, Menlo Park, CA, 94025, USA}
\author{L.~E.~Bleem\,\orcidlink{0000-0001-7665-5079}}
\affiliation{High-Energy Physics Division, Argonne National Laboratory, 9700 South Cass Avenue., Lemont, IL, 60439, USA}
\affiliation{Kavli Institute for Cosmological Physics, University of Chicago, 5640 South Ellis Avenue, Chicago, IL, 60637, USA}
\author{F.~R.~Bouchet\,\orcidlink{0000-0002-8051-2924}}
\affiliation{Institut d'Astrophysique de Paris, UMR 7095, CNRS \& Sorbonne Universit\'{e}, 98 bis boulevard Arago, 75014 Paris, France}
\author{L.~Bryant}
\affiliation{Enrico Fermi Institute, University of Chicago, 5640 South Ellis Avenue, Chicago, IL, 60637, USA}
\author{E.~Camphuis}
\affiliation{Institut d'Astrophysique de Paris, UMR 7095, CNRS \& Sorbonne Universit\'{e}, 98 bis boulevard Arago, 75014 Paris, France}
\author{J.~E.~Carlstrom}
\affiliation{Kavli Institute for Cosmological Physics, University of Chicago, 5640 South Ellis Avenue, Chicago, IL, 60637, USA}
\affiliation{Enrico Fermi Institute, University of Chicago, 5640 South Ellis Avenue, Chicago, IL, 60637, USA}
\affiliation{Department of Physics, University of Chicago, 5640 South Ellis Avenue, Chicago, IL, 60637, USA}
\affiliation{High-Energy Physics Division, Argonne National Laboratory, 9700 South Cass Avenue., Lemont, IL, 60439, USA}
\affiliation{Department of Astronomy and Astrophysics, University of Chicago, 5640 South Ellis Avenue, Chicago, IL, 60637, USA}
\author{T.~W.~Cecil\,\orcidlink{0000-0002-7019-5056}}
\affiliation{High-Energy Physics Division, Argonne National Laboratory, 9700 South Cass Avenue., Lemont, IL, 60439, USA}
\author{C.~L.~Chang}
\affiliation{High-Energy Physics Division, Argonne National Laboratory, 9700 South Cass Avenue., Lemont, IL, 60439, USA}
\affiliation{Kavli Institute for Cosmological Physics, University of Chicago, 5640 South Ellis Avenue, Chicago, IL, 60637, USA}
\affiliation{Department of Astronomy and Astrophysics, University of Chicago, 5640 South Ellis Avenue, Chicago, IL, 60637, USA}
\author{P.~Chaubal}
\affiliation{School of Physics, University of Melbourne, Parkville, VIC 3010, Australia}
\author{P.~M.~Chichura\,\orcidlink{0000-0002-5397-9035}}
\affiliation{Department of Physics, University of Chicago, 5640 South Ellis Avenue, Chicago, IL, 60637, USA}
\affiliation{Kavli Institute for Cosmological Physics, University of Chicago, 5640 South Ellis Avenue, Chicago, IL, 60637, USA}
\author{T.-L.~Chou}
\affiliation{Department of Physics, University of Chicago, 5640 South Ellis Avenue, Chicago, IL, 60637, USA}
\affiliation{Kavli Institute for Cosmological Physics, University of Chicago, 5640 South Ellis Avenue, Chicago, IL, 60637, USA}
\author{T.~M.~Crawford\,\orcidlink{0000-0001-9000-5013}}
\affiliation{Kavli Institute for Cosmological Physics, University of Chicago, 5640 South Ellis Avenue, Chicago, IL, 60637, USA}
\affiliation{Department of Astronomy and Astrophysics, University of Chicago, 5640 South Ellis Avenue, Chicago, IL, 60637, USA}
\author{A.~Cukierman}
\affiliation{Kavli Institute for Particle Astrophysics and Cosmology, Stanford University, 452 Lomita Mall, Stanford, CA, 94305, USA}
\affiliation{SLAC National Accelerator Laboratory, 2575 Sand Hill Road, Menlo Park, CA, 94025, USA}
\affiliation{Department of Physics, Stanford University, 382 Via Pueblo Mall, Stanford, CA, 94305, USA}
\author{C.~Daley\,\orcidlink{0000-0002-3760-2086}}
\affiliation{Department of Astronomy, University of Illinois Urbana-Champaign, 1002 West Green Street, Urbana, IL, 61801, USA}
\author{T.~de~Haan}
\affiliation{High Energy Accelerator Research Organization (KEK), Tsukuba, Ibaraki 305-0801, Japan}
\author{K.~R.~Dibert}
\affiliation{Department of Astronomy and Astrophysics, University of Chicago, 5640 South Ellis Avenue, Chicago, IL, 60637, USA}
\affiliation{Kavli Institute for Cosmological Physics, University of Chicago, 5640 South Ellis Avenue, Chicago, IL, 60637, USA}
\author{M.~A.~Dobbs}
\affiliation{Department of Physics and McGill Space Institute, McGill University, 3600 Rue University, Montreal, Quebec H3A 2T8, Canada}
\affiliation{Canadian Institute for Advanced Research, CIFAR Program in Gravity and the Extreme Universe, Toronto, ON, M5G 1Z8, Canada}
\author{A.~Doussot}
\affiliation{Institut d'Astrophysique de Paris, UMR 7095, CNRS \& Sorbonne Universit\'{e}, 98 bis boulevard Arago, 75014 Paris, France}
\author{D.~Dutcher\,\orcidlink{0000-0002-9962-2058}}
\affiliation{Joseph Henry Laboratories of Physics, Jadwin Hall, Princeton University, Princeton, NJ 08544, USA}
\author{W.~Everett}
\affiliation{Department of Astrophysical and Planetary Sciences, University of Colorado, Boulder, CO, 80309, USA}
\author{C.~Feng}
\affiliation{Department of Physics, University of Illinois Urbana-Champaign, 1110 West Green Street, Urbana, IL, 61801, USA}
\author{A.~Foster\,\orcidlink{0000-0002-7145-1824}}
\affiliation{Department of Physics, Case Western Reserve University, Cleveland, OH, 44106, USA}
\author{S.~Galli}
\affiliation{Institut d'Astrophysique de Paris, UMR 7095, CNRS \& Sorbonne Universit\'{e}, 98 bis boulevard Arago, 75014 Paris, France}
\author{A.~E.~Gambrel}
\affiliation{Kavli Institute for Cosmological Physics, University of Chicago, 5640 South Ellis Avenue, Chicago, IL, 60637, USA}
\author{R.~W.~Gardner}
\affiliation{Enrico Fermi Institute, University of Chicago, 5640 South Ellis Avenue, Chicago, IL, 60637, USA}
\author{N.~Goeckner-Wald}
\affiliation{Department of Physics, Stanford University, 382 Via Pueblo Mall, Stanford, CA, 94305, USA}
\affiliation{Kavli Institute for Particle Astrophysics and Cosmology, Stanford University, 452 Lomita Mall, Stanford, CA, 94305, USA}
\author{R.~Gualtieri\,\orcidlink{0000-0003-4245-2315}}
\affiliation{High-Energy Physics Division, Argonne National Laboratory, 9700 South Cass Avenue., Lemont, IL, 60439, USA}
\author{F.~Guidi}
\affiliation{Institut d'Astrophysique de Paris, UMR 7095, CNRS \& Sorbonne Universit\'{e}, 98 bis boulevard Arago, 75014 Paris, France}
\author{S.~Guns}
\affiliation{Department of Physics, University of California, Berkeley, CA, 94720, USA}
\author{N.~W.~Halverson}
\affiliation{CASA, Department of Astrophysical and Planetary Sciences, University of Colorado, Boulder, CO, 80309, USA }
\affiliation{Department of Physics, University of Colorado, Boulder, CO, 80309, USA}
\author{E.~Hivon}
\affiliation{Institut d'Astrophysique de Paris, UMR 7095, CNRS \& Sorbonne Universit\'{e}, 98 bis boulevard Arago, 75014 Paris, France}
\author{G.~P.~Holder\,\orcidlink{0000-0002-0463-6394}}
\affiliation{Department of Physics, University of Illinois Urbana-Champaign, 1110 West Green Street, Urbana, IL, 61801, USA}
\author{W.~L.~Holzapfel}
\affiliation{Department of Physics, University of California, Berkeley, CA, 94720, USA}
\author{J.~C.~Hood}
\affiliation{Kavli Institute for Cosmological Physics, University of Chicago, 5640 South Ellis Avenue, Chicago, IL, 60637, USA}
\author{N.~Huang}
\affiliation{Department of Physics, University of California, Berkeley, CA, 94720, USA}
\author{L.~Knox}
\affiliation{Department of Physics \& Astronomy, University of California, One Shields Avenue, Davis, CA 95616, USA}
\author{M.~Korman}
\affiliation{Department of Physics, Case Western Reserve University, Cleveland, OH, 44106, USA}
\author{C.-L.~Kuo}
\affiliation{Kavli Institute for Particle Astrophysics and Cosmology, Stanford University, 452 Lomita Mall, Stanford, CA, 94305, USA}
\affiliation{Department of Physics, Stanford University, 382 Via Pueblo Mall, Stanford, CA, 94305, USA}
\affiliation{SLAC National Accelerator Laboratory, 2575 Sand Hill Road, Menlo Park, CA, 94025, USA}
\author{A.~T.~Lee}
\affiliation{Department of Physics, University of California, Berkeley, CA, 94720, USA}
\affiliation{Physics Division, Lawrence Berkeley National Laboratory, Berkeley, CA, 94720, USA}
\author{A.~E.~Lowitz}
\affiliation{Kavli Institute for Cosmological Physics, University of Chicago, 5640 South Ellis Avenue, Chicago, IL, 60637, USA}
\author{C.~Lu}
\affiliation{Department of Physics, University of Illinois Urbana-Champaign, 1110 West Green Street, Urbana, IL, 61801, USA}
\author{M.~Millea\,\orcidlink{0000-0001-7317-0551}}
\affiliation{Department of Physics, University of California, Berkeley, CA, 94720, USA}
\author{J.~Montgomery}
\affiliation{Department of Physics and McGill Space Institute, McGill University, 3600 Rue University, Montreal, Quebec H3A 2T8, Canada}
\author{T.~Natoli}
\affiliation{Kavli Institute for Cosmological Physics, University of Chicago, 5640 South Ellis Avenue, Chicago, IL, 60637, USA}
\author{G.~I.~Noble}
\affiliation{Department of Physics and McGill Space Institute, McGill University, 3600 Rue University, Montreal, Quebec H3A 2T8, Canada}
\author{V.~Novosad}
\affiliation{Materials Sciences Division, Argonne National Laboratory, 9700 South Cass Avenue, Lemont, IL, 60439, USA}
\author{Y.~Omori}
\affiliation{Department of Astronomy and Astrophysics, University of Chicago, 5640 South Ellis Avenue, Chicago, IL, 60637, USA}
\affiliation{Kavli Institute for Cosmological Physics, University of Chicago, 5640 South Ellis Avenue, Chicago, IL, 60637, USA}
\author{S.~Padin}
\affiliation{Kavli Institute for Cosmological Physics, University of Chicago, 5640 South Ellis Avenue, Chicago, IL, 60637, USA}
\affiliation{California Institute of Technology, 1200 East California Boulevard., Pasadena, CA, 91125, USA}
\author{Z.~Pan}
\affiliation{High-Energy Physics Division, Argonne National Laboratory, 9700 South Cass Avenue., Lemont, IL, 60439, USA}
\affiliation{Kavli Institute for Cosmological Physics, University of Chicago, 5640 South Ellis Avenue, Chicago, IL, 60637, USA}
\affiliation{Department of Physics, University of Chicago, 5640 South Ellis Avenue, Chicago, IL, 60637, USA}
\author{P.~Paschos}
\affiliation{Enrico Fermi Institute, University of Chicago, 5640 South Ellis Avenue, Chicago, IL, 60637, USA}
\author{K.~Prabhu}
\affiliation{Department of Physics \& Astronomy, University of California, One Shields Avenue, Davis, CA 95616, USA}
\author{W.~Quan}
\affiliation{Department of Physics, University of Chicago, 5640 South Ellis Avenue, Chicago, IL, 60637, USA}
\affiliation{Kavli Institute for Cosmological Physics, University of Chicago, 5640 South Ellis Avenue, Chicago, IL, 60637, USA}
\author{A.~Rahlin\,\orcidlink{0000-0003-3953-1776}}
\affiliation{Fermi National Accelerator Laboratory, MS209, P.O. Box 500, Batavia, IL, 60510, USA}
\affiliation{Kavli Institute for Cosmological Physics, University of Chicago, 5640 South Ellis Avenue, Chicago, IL, 60637, USA}
\author{C.~L.~Reichardt\,\orcidlink{0000-0003-2226-9169}}
\affiliation{School of Physics, University of Melbourne, Parkville, VIC 3010, Australia}
\author{M.~Rouble}
\affiliation{Department of Physics and McGill Space Institute, McGill University, 3600 Rue University, Montreal, Quebec H3A 2T8, Canada}
\author{J.~E.~Ruhl}
\affiliation{Department of Physics, Case Western Reserve University, Cleveland, OH, 44106, USA}
\author{E.~Schiappucci}
\affiliation{School of Physics, University of Melbourne, Parkville, VIC 3010, Australia}
\author{G.~Smecher}
\affiliation{Three-Speed Logic, Inc., Victoria, B.C., V8S 3Z5, Canada}
\author{J.~A.~Sobrin}
\affiliation{Department of Physics, University of Chicago, 5640 South Ellis Avenue, Chicago, IL, 60637, USA}
\affiliation{Kavli Institute for Cosmological Physics, University of Chicago, 5640 South Ellis Avenue, Chicago, IL, 60637, USA}
\author{J.~Stephen}
\affiliation{Enrico Fermi Institute, University of Chicago, 5640 South Ellis Avenue, Chicago, IL, 60637, USA}
\author{A.~Suzuki}
\affiliation{Physics Division, Lawrence Berkeley National Laboratory, Berkeley, CA, 94720, USA}
\author{C.~Tandoi}
\affiliation{Department of Astronomy, University of Illinois Urbana-Champaign, 1002 West Green Street, Urbana, IL, 61801, USA}
\author{K.~L.~Thompson}
\affiliation{Kavli Institute for Particle Astrophysics and Cosmology, Stanford University, 452 Lomita Mall, Stanford, CA, 94305, USA}
\affiliation{Department of Physics, Stanford University, 382 Via Pueblo Mall, Stanford, CA, 94305, USA}
\affiliation{SLAC National Accelerator Laboratory, 2575 Sand Hill Road, Menlo Park, CA, 94025, USA}
\author{B.~Thorne}
\affiliation{Department of Physics \& Astronomy, University of California, One Shields Avenue, Davis, CA 95616, USA}
\author{C.~Tucker}
\affiliation{School of Physics and Astronomy, Cardiff University, Cardiff CF24 3YB, United Kingdom}
\author{C.~Umilta\,\orcidlink{0000-0002-6805-6188}}
\affiliation{Department of Physics, University of Illinois Urbana-Champaign, 1110 West Green Street, Urbana, IL, 61801, USA}
\author{J.~D.~Vieira}
\affiliation{Department of Astronomy, University of Illinois Urbana-Champaign, 1002 West Green Street, Urbana, IL, 61801, USA}
\affiliation{Department of Physics, University of Illinois Urbana-Champaign, 1110 West Green Street, Urbana, IL, 61801, USA}
\affiliation{Center for AstroPhysical Surveys, National Center for Supercomputing Applications, Urbana, IL, 61801, USA}
\author{G.~Wang}
\affiliation{High-Energy Physics Division, Argonne National Laboratory, 9700 South Cass Avenue., Lemont, IL, 60439, USA}
\author{W.~L.~K.~Wu\,\orcidlink{0000-0001-5411-6920}}
\affiliation{Kavli Institute for Particle Astrophysics and Cosmology, Stanford University, 452 Lomita Mall, Stanford, CA, 94305, USA}
\affiliation{SLAC National Accelerator Laboratory, 2575 Sand Hill Road, Menlo Park, CA, 94025, USA}
\author{V.~Yefremenko}
\affiliation{High-Energy Physics Division, Argonne National Laboratory, 9700 South Cass Avenue., Lemont, IL, 60439, USA}
\author{M.~R.~Young}
\affiliation{David A. Dunlap Department of Astronomy \& Astrophysics, University of Toronto, 50 St. George Street, Toronto, ON, M5S 3H4, Canada}
\collaboration{SPT-3G Collaboration}
\noaffiliation

\begin{abstract}
    Ultralight axionlike particles (ALPs) are compelling dark matter candidates because of their potential to resolve small-scale discrepancies between $\Lambda$CDM predictions and cosmological observations. Axion-photon coupling induces a polarization rotation in linearly polarized photons traveling through an ALP field; thus, as the local ALP dark matter field oscillates in time, distant static polarized sources will appear to oscillate with a frequency proportional to the ALP mass. We use observations of the cosmic microwave background from SPT-3G, the current receiver on the South Pole Telescope, to set upper limits on the value of the axion-photon coupling constant $g_{\phi\gamma}$ over the approximate mass range $10^{-22} - 10^{-19}$ eV, corresponding to oscillation periods from $12$ hours to $100$ days. For periods between $1$ and $100$ days ($4.7 \times 10^{-22} \text{ eV} \leq m_\phi \leq 4.7 \times 10^{-20} \text{ eV}$), where the limit is approximately constant, we set a median $95\%$ C.L. upper limit on the amplitude of on-sky polarization rotation of $0.071$ deg. Assuming that dark matter comprises a single ALP species with a local dark matter density of $0.3$ GeV/cm$^3$, this corresponds to $g_{\phi\gamma} < 1.18 \times 10^{-12}\text{ GeV}^{-1} \times \left( \frac{m_{\phi}}{1.0 \times 10^{-21} \text{ eV}} \right)$. These new limits represent an improvement over the previous strongest limits set using the same effect by a factor of ${\sim} 3.8$.
\end{abstract}


\maketitle

\section{Introduction}
\label{sec:intro}

Astrophysical observations have provided strong evidence for the existence of nonbaryonic dark matter \citep{persic96, garrett11}. The QCD axion, originally devised to solve the strong $CP$ problem \citep{weinberg78, wilczek78, peccei77a, peccei77b}, has emerged as a compelling dark matter candidate \citep{preskill83, abbott83, dine83, duffy09, graham15}, although theoretical considerations constrain the region of mass parameter space it can lie in. Of broader astrophysical interest is a class of axionlike particles (ALPs) that arise naturally in many string theory models \citep{witten84, arvanitaki10, cicoli12}. Although they couple to the Standard Model photon in much the same way as the QCD axion, ALPs do not solve the strong $CP$ problem. Despite this, they make promising dark matter candidates, as they may lie in a much wider region of parameter space than the highly constrained QCD axion \citep{frieman95, amendola06}. For convenience, we will use ``axion'' as an umbrella term encompassing both the QCD axion and ALPs.\blankfootnote{*Corresponding author.}\blankfootnote{kferguson@physics.ucla.edu}

Many experiments have carried out axion searches. Generally these searches take advantage of the coupling between axions and photons via the Primakoff effect, by which an axion is converted into a photon (or vice versa) in the presence of a strong magnetic field. Helioscope experiments such as CAST \citep{cast17} are able to set limits on the axion-photon coupling constant $g_{\phi\gamma}$ across a wide range of possible axion masses $m_{\phi}$, with the upper mass range given by instrumental considerations rather than a theoretical limit. Haloscopes like ADMX \citep{admx21} and HAYSTAC \citep{zhong18} instead use resonant cavities to set stringent limits on $g_{\phi\gamma}$ in narrow windows of mass within the favored range of masses for the QCD axion.

The axion contributes an additional term to the photon equations of motion in the form of an imaginary exponential. The consequence of this is that opposite-helicity photons pick up relative phase shifts as they travel through an axion field \citep[][hereafter F19]{fedderke19}. From the point of view of an observer, the polarization angle of a linearly polarized photon will be rotated by an amount proportional to the difference between the axion field values at emission and absorption. Searches for this effect often focus on ultralight axions (those with masses roughly between $10^{-23}$ eV and $10^{-18}$ eV) because cold axions with these masses form a Bose-Einstein condensate and thus behave as a classical field with a value that oscillates on human-observable timescales, with periods in the range from hours to years. Additionally, ultralight axions are especially interesting as a dark matter candidate due to the long de Broglie wavelengths of their condensate fields; their scale-dependent clustering has the potential to resolve long-standing discrepancies between observations and predictions of the standard cosmological model $\Lambda$CDM on small scales, such as the core/cusp problem and the too-big-to-fail problem \citep{hu00c, ferreira21}. Because thermally produced axions in this mass range would still be relativistic today, it is important that they be produced nonthermally for them to remain a viable dark matter candidate. This may happen via vacuum realignment, string decay, or domain wall decay \citep{sikivie08, duffy09}.

Using active galactic nuclei (AGN) as astrophysical polarization sources, \citet{horns12} and \citet{ivanov19} set limits on $g_{\phi\gamma}$ for ultralight axions. However, intrinsic variation in the polarization of AGN sources can be difficult to disentangle from an axion signal; along with uncertainty in the dark matter density at the source and uncertainty in modeling the magnetic field around the AGN, there are major systematics that must be accounted for. These difficulties are somewhat alleviated by using galactic pulsars as astrophysical polarization sources, as in \citet{castillo22}. Interferometric laboratory searches utilizing this polarization-rotation effect, such as DANCE \citep{michimura20} and ADAM-GD \citep{nagano21}, promise significant increases in sensitivity over the current state of the art at a wide range of masses, but such searches are in the early stages with results still many years away.

F19 proposed using the cosmic microwave background (CMB) as a source with which to carry out an axion search. Searches using the CMB have smaller systematic uncertainties than those using AGN because the polarization of the CMB has no intrinsic time variation on the experiment-relevant scales of hours to years. Compared to future laboratory searches for time-dependent birefringence, CMB experiments have datasets currently available that span many years and cover significant fractions of the sky. The noise properties of these datasets are sufficiently well understood to measure time-varying birefringence across an interesting range of $g_{\phi\gamma}$.

Ultralight axions have two main effects on CMB measurements. The first effect (what F19 call the \textit{washout effect}) accounts for the fact that the CMB was not formed instantaneously, but rather photons decoupled over the course of ${\sim} 100{,}000$ years. In the mass range considered in this work (approximately $10^{-22}$ eV to $10^{-19}$ eV, corresponding to oscillations on the order of hours to years), the axion field oscillates many times over the visibility function of the CMB at last scattering. This leads to an averaging effect which causes the CMB we observe today to have a slightly reduced polarization that is static in time, manifesting as a slight suppression of the CMB polarization power spectra. Second, in what F19 call the \textit{AC oscillation effect}, the oscillation of the local axion dark matter field induces a time-dependent birefringence effect, causing the polarization angle of CMB photons to oscillate in time. Because the coherence length of the local axion field is so large at the masses under consideration, this oscillation is coherent over long periods of time. Additionally, because the measured rotation is set by the local value of the axion field, the oscillation appears in phase across the entire sky. CMB experiments can measure the amount of polarization rotation as a function of time, directly measuring the effect of the dark matter. Constraints from the washout effect are fundamentally limited due to cosmic variance (that is, the fundamental statistical uncertainty or sample variance that arises due to the fact that there are a finite number of modes a CMB experiment could observe from our fixed location relative to the CMB), with the current constraints a factor of ${\sim} \sqrt{7}$ away from this limit \citep{fedderke19}. Therefore future discovery potential must rely on the AC oscillation effect. The BICEP/Keck collaboration has recently published results of searches for this AC oscillation effect, demonstrating its viability as a search technique \citep[][hereafter BK22]{bicepkeck21, bicepkeck22}.

In this paper, we describe a search for the AC oscillation effect using SPT-3G, the current camera installed on the South Pole Telescope (SPT), in which we measure a time series of polarization rotation angles and associated uncertainties, fit a sinusoidal model, and extract limits on $g_{\phi\gamma}$.  We set the tightest limits on axion dark matter through the AC oscillation effect to date, improving on current limits by a factor of ${\sim} 3.8$ and approximately matching the limit from the washout effect. In Sec. \ref{sec:instrument}, SPT-3G is described, with particular attention paid towards why it is an ideal instrument with which to carry out this search. In Sec. \ref{sec:methods}, the details of the analysis procedure are laid out. Results and discussion of the broader context follow in Sec. \ref{sec:results}.


\section{Instrument and Dataset}
\label{sec:instrument}

The SPT is a $10$-meter millimeter-wavelength telescope located at the Amundsen-Scott South Pole Station in Antarctica \citep{carlstrom11}. The current camera installed on the telescope is SPT-3G, an array of ${\sim} 16{,}000$ polarization-sensitive transition-edge sensor (TES) bolometers \citep{sobrin22}. As detailed in \citet{sobrin22}, the bolometers are cooled to an operating temperature of $300$ mK by a 3He/4He sorption cooler for ${\sim} 15$ hours at a time, separated by a ${\sim} 4.5$ hour interval when the cooler is re-cycled. SPT-3G is designed to observe the CMB in three bands, centered at approximately $95$, $150$, and $220$ GHz, with an angular resolution of approximately $1.2$ arcminutes at $150$ GHz.

In an ongoing multiyear survey, SPT-3G is used to observe a ${\sim} 1500 \deg^2$ patch of the sky spanning $-50\deg$ to $50\deg$ in right ascension (RA) and $-70\deg$ to $-42\deg$ in declination. The full survey field is broken up into four subfields, each spanning the full range in RA and centered on $-44.75\deg, -52.25\deg, -59.75\deg$, and $-67.25\deg$ in declination. In a subset of data called a \textit{scan}, the telescope sweeps across the entire RA range at a constant velocity and elevation (corresponding to a nearly constant declination due to its location roughly a kilometer from the geographical South Pole). The telescope performs two scans in opposite directions (a \textit{scan pair}) at the same elevation before stepping up $12.5$ arcminutes; this process is then repeated until the entire declination range of a subfield has been covered. The combination of all scans together is called an \textit{observation}. Each observation takes approximately two hours and generates a set of time-ordered data (TOD) for each bolometer that can later be turned into maps of the sky (Sec. \ref{sec:TOD-maps}). In addition to the survey field observations, SPT-3G also takes regular calibration observations, which are described in more detail in \citet{sobrin22} and \citet{dutcher21} (hereafter D21).

For the work presented here, we use data from SPT-3G's $2019$ observing season. Specifically, we use only the $95$ GHz and $150$ GHz bands, as they have the highest CMB sensitivity. Gaps between the panels of the telescope primary mirror create diffraction sidelobes, which can couple to the sun and produce stripes in the SPT-3G maps. To avoid this systematic signal, we limit ourselves to data between March $22$, $2019$ (sunset at the South Pole) and November $30$, $2019$. These choices are conservative cuts motivated by an internal analysis examining the time dependence of sun contamination in the maps. As part of our suite of jackknife tests (detailed in Sec. \ref{sec:jackknives}), we also test the remaining data for evidence of sun contamination.

SPT-3G is well suited to perform a search for the AC oscillation effect. Its location at the South Pole allows it to observe the same patch of sky regardless of the rotation of the earth. The combination of a long period of observation with finely sampled individual observations allows it to be sensitive to oscillation frequencies (and therefore axion masses) spanning more than three orders of magnitude. Finally, due to its high angular resolution, SPT can measure the CMB E-mode power with S/N $\gtrsim 1$ to small angular scales. In particular, SPT is sensitive to ${\sim} 16$ times as many modes as BK22 (which has an angular resolution of ${\sim} 0.5$ deg at $150$ GHz), allowing it to set tighter limits than BK22 on $g_{\phi\gamma}$ by a factor of ${\sim} 4$ [see Eq. (\ref{eqn:sensitivity-model})].


\section{Methods}
\label{sec:methods}
The analysis proceeds as follows: maps of each observation are created from the TOD (Sec. \ref{sec:TOD-maps}); particularly noisy maps are cut (Sec. \ref{sec:datacuts}); for each observation, we calculate a polarization rotation angle and uncertainty (Sec. \ref{sec:maps-angs}); we analyze the resulting time series of angles for systematic effects (Sec. \ref{sec:jackknives}); we then search for a periodic signal in this time series (Sec. \ref{sec:angs-lims}).

\subsection{Time-ordered data to maps}
\label{sec:TOD-maps}
The raw TOD from each scan are converted into CMB temperature units, filtered, and binned into maps in the manner described in D21, giving us the intermediate data products of one map per scan. We can then \textit{coadd} (that is, perform a weighted average of) the per-scan maps into a single map per observation. There are three differences between D21 and the current work:

\begin{itemize}
\item[(i)] To reduce the amount of aliased power in the maps, we set the cutoff for the low-pass TOD filter at $\ell=5000$ rather than $\ell=6600$.
\item[(ii)] The source list used for masking/interpolating during TOD filtering comprises all sources detected in $2018$ data with a signal-to-noise ratio of greater than $10$ in the $95$ GHz observing band.
\item[(iii)] Lastly, although we only calculate polarization rotation angles on coadded single-observation maps, we choose to save maps of every individual scan rather than coadded left- or right-going maps as in D21. This allows a more detailed understanding of the statistical properties of individual observations, which provides valuable information when deciding which observations to cut. Additionally, it allows us to generate many noise realizations per observation, which is necessary to determine the uncertainty of the per-observation polarization rotation angle (see Sec. \ref{sec:uncertainty} for details).
\end{itemize}

Map-space weights are also calculated in this step. We first calculate the power spectral density of each \textit{timestream} (that is, the TOD for a single bolometer for a single scan) and determine the variance in the timestream by integrating the power between $1.0$ Hz and $4.0$ Hz. The timestreams are inverse-variance weighted, and the weights in map space are the sum of the weights of the specific bolometers that are binned into each pixel (see D21 for further details). These weights are used to determine the data quality in an observation (Sec. \ref{sec:datacuts}) and coadd individual observation maps into a full season map (Sec. \ref{sec:coadds}).

\subsection{Data cuts}
\label{sec:datacuts}
In order to prevent particularly noisy or miscalibrated timestreams from being coadded into maps, individual detectors are flagged and their TOD cut during every scan. As in D21, leading reasons detectors may be flagged are: having anomalous calibration statistics; dropping out of the superconducting transition; having too large a variance in the timestream; or being subject to large, sudden shifts (denoted \textit{glitches}) in their timestreams. The only difference is that significant improvements were made to the glitch-finding algorithm between D21 and the current work. On average per scan, in the $95$ ($150$) GHz band, we flag $1091$ ($925$) bolometer timestreams, which leaves $3489$ ($3641$) bolometer timestreams that are binned into the maps.

Even after flagging bad bolometer timestreams, some single-observation maps will have undesirable noise properties; for this reason, we institute additional cuts on entire maps (choosing cutoff thresholds so as to cut any clear outliers). We implement a few cuts based on the map weights: observations with median weights below a cutoff threshold are cut due to their high noise level; observations with median weights above another cutoff threshold are also cut on the basis that they are unphysical. We also want to cut observations with nonuniform weights, as this usually indicates a significant change in weather or detector responsivity over the course of the observation. To identify these observations, we calculate the standard deviation of the weights divided by the median weight for each observation, cutting any where this quantity is above a cutoff threshold. We cut all maps for observations that were aborted early, as this usually signals an early end to the fridge cycle and thus it is assumed that the data before the observation was stopped are tainted by degraded cryogenic performance. Finally, we construct simulated maps (see Sec. \ref{sec:coadds} for details) with opposite-direction scans subtracted from, rather than coadded to, each other. The polarization rotation angles computed from these maps should be consistent with zero; thus as a final cut, we flag any observation where either this angle or the angle divided by its uncertainty is above a cutoff threshold.

SPT-3G took $1604$ observations split across the four subfields between our chosen start and end dates. With the chosen cutoff values, we flag $59$ observations for cutting, amounting to a $3.7\%$ reduction in data volume.

\subsection{Maps to angles}
\label{sec:maps-angs}
Once maps have been made, we calculate the magnitude of the on-sky polarization rotation angle for each observation for each observing band. In terms of quantities that we measure with SPT-3G, the polarization rotation manifests as a rotation of the Stokes $Q$ parameter into Stokes $U$ (and vice versa). These maps include polarized signals from both the CMB and astrophysical foregrounds; while the rotation of the foregrounds is not necessarily in phase with that of the CMB, the CMB signal is dominant over the foreground signal in the SPT-3G patch of the sky. Thus, it is a fair assumption that any observed time-dependent birefringence would be dominated by the rotation of polarized CMB photons. In the limit of a small rotation amplitude, our model for the measured $Q$ and $U$ is

\begin{align}
\label{eqn:model}
\begin{split}
Q^{\textrm{m}}_i \left(\rho\right) &= Q_{0,i} - \rho U_{0,i}, \\
U^{\textrm{m}}_i \left(\rho\right) &= U_{0,i} + \rho Q_{0,i},
\end{split}
\end{align}
where the ``$\textrm{m}$'' superscript denotes \textit{model}, the $0$ subscript denotes the $Q$ and $U$ fields that would be measured in the limit where $g_{\phi\gamma} = 0$, $i$ represents the index of an individual map pixel (since the rotation is the same across the entire map), and $\rho$ is the polarization rotation angle induced by the axion.\footnote{The true on-sky rotation angle $\rho_\textrm{sky}$ is related to the $Q$/$U$ rotation angle by a factor of $2$: $\rho_\textrm{sky} = \rho / 2$.} We model $\rho$ as a function of time $t$,

\begin{align}
\label{eqn:angle-model}
\begin{split}
\rho_\textrm{m} \left(t\right) &= A \sin\left( 2\pi f t + \delta \right) \\
                   &= g_{\phi\gamma}\phi_{0} \sin\left( m_{\phi} t + \delta \right),
\end{split}
\end{align}
where $A$ is the amplitude of the oscillation, $f$ is its frequency, $\delta$ is the phase, $\phi_{0}$ is the maximum value of the local axion field, and $m_{\phi}$ is the axion mass.

We do not know the true CMB fields $Q_0$ and $U_0$, so we use the full-season coadded and filtered $Q$ and $U$ maps as estimates (further details in Sec. \ref{sec:coadds}). As a consequence of this choice, all single-observation angles $\rho$ are measured relative to the season-long average. For low-frequency modes, this has the effect of reducing the constraining power of our limits, though due to the ${\sim} 250$-day span of our data, the effect is negligible for even the lowest frequency we consider ($0.01$ inverse-days). Additionally, this means that by construction we do not measure any DC rotation (that is, any constant birefringence).

In order to estimate $\rho$, we coadd our individual-scan maps into a single complete-observation map. We then construct the map-space quantity

\begin{equation}
\label{eqn:chi2}
\chi^2 \left(\rho\right) = \sum_{pq,ij} \left( P_{pi} - P^{\textrm{m}}_{pi} \left(\rho\right) \right) \left( \mathbf{C}^{-1} \right)_{pq,ij}  \left( P_{qj} - P^{\textrm{m}}_{qj} \left(\rho\right) \right),
\end{equation}
where $P_{pi}$ represents the observed $Q$ and $U$ maps at pixel $i$ (i.e., $p\in\{ Q,U \}$ with $P_{Qi} = Q_i$ and $P_{Ui} = U_i$), $P^\textrm{m}_{pi}$ represents the model expectation for Stokes parameter $p$ at pixel $i$ [given by Eq. (\ref{eqn:model})], and $\mathbf{C}_{pq,ij}$ is the map-domain covariance between all pixels and $Q$ and $U$ maps.

The best-fit rotation angle $\hat{\rho}$ is determined by minimizing the $\chi^2$ with respect to $\rho$. We can derive an analytical expression for $\hat{\rho}$ if we assume that the covariance $\mathbf{C}_{pq,ij}$ is diagonal in ${i,j}$; that is, that there is no pixel-pixel covariance. For maps with our chosen $2$-arcminute resolution, the average pixel-pixel covariance in $Q$ and $U$ maps is negligible for all but a pixel's nearest neighbors, where it is approximately at the $10\%$ level. Neglecting this covariance causes Eq. (\ref{eqn:chi2}) to be slightly non-$\chi^2$ distributed. While this means we cannot use its asymptotic form for hypothesis tests, this is not strictly necessary and so we choose to neglect the covariance here; it is instead implicitly included in the process for determining the uncertainty on $\hat{\rho}$ (Sec. \ref{sec:uncertainty}). Thus we set $\left( \mathbf{C}^{-1} \right)_{pq,ij} = 0$ for all $i \neq j$. Because our maps are inverse-variance weighted, we can replace this quantity with the polarization weight matrix $\mathbf{W}$ (that is, $\mathbf{C}_{pq,ii} = 1/\mathbf{W}_{pq,i}$). Writing all terms out in explicit detail, we determine that

\begin{widetext}
\begin{equation}
\hat{\rho} = \frac{\sum_i \mathbf{W}_{QQ,i} \left( Q_{0,i} U_{0,i} - Q_i U_{0,i} \right) + \mathbf{W}_{UU,i} \left( Q_{0,i}U_i - Q_{0,i} U_{0,i} \right) + \mathbf{W}_{QU,i} \left( Q_i Q_{0,i} - U_i U_{0,i} - Q_{0,i}^2 + U_{0,i}^2 \right)}{\sum_i \mathbf{W}_{QQ,i} U_{0,i}^2 + \mathbf{W}_{UU,i} Q_{0,i}^2 - 2\mathbf{W}_{QU,i} Q_{0,i}U_{0,i}},
\end{equation}
\end{widetext}
where the sum over $i$ is a sum over the pixels in the map.

Because each observation takes ${\sim} 2$ hours, we cannot instantaneously sample the polarization rotation angle $\rho$. We assume then that our estimated angle $\hat{\rho}$ is actually an average over the true signal,

\begin{align}
\label{eqn:average}
\begin{split}
\hat{\rho} &= \frac{1}{t_2 - t_1} \int_{t_1}^{t_2} A \sin\left( 2\pi f t + \delta \right) dt \\
           &= \rho_\textrm{m}\left( \tau \right) \sinc\left[\pi f \left( t_2 - t_1 \right)\right] \\
           &= \rho_\textrm{m}\left( \tau \right) \sinc\left( \frac{m_\phi \left( t_2 - t_1 \right)}{2} \right),
\end{split}
\end{align}
where $\tau$ is the mean time of the observation $\frac{t_1 + t_2}{2}$ and we use the unnormalized $\sinc$ function. The effect of this averaging is mostly negligible; our sensitivity is reduced by only ${\sim} 5\%$ at even the highest frequency we consider ($2.0$ inverse-days).

\label{sec:coadds}
\begin{figure*}
\includegraphics[width=1.5\columnwidth]{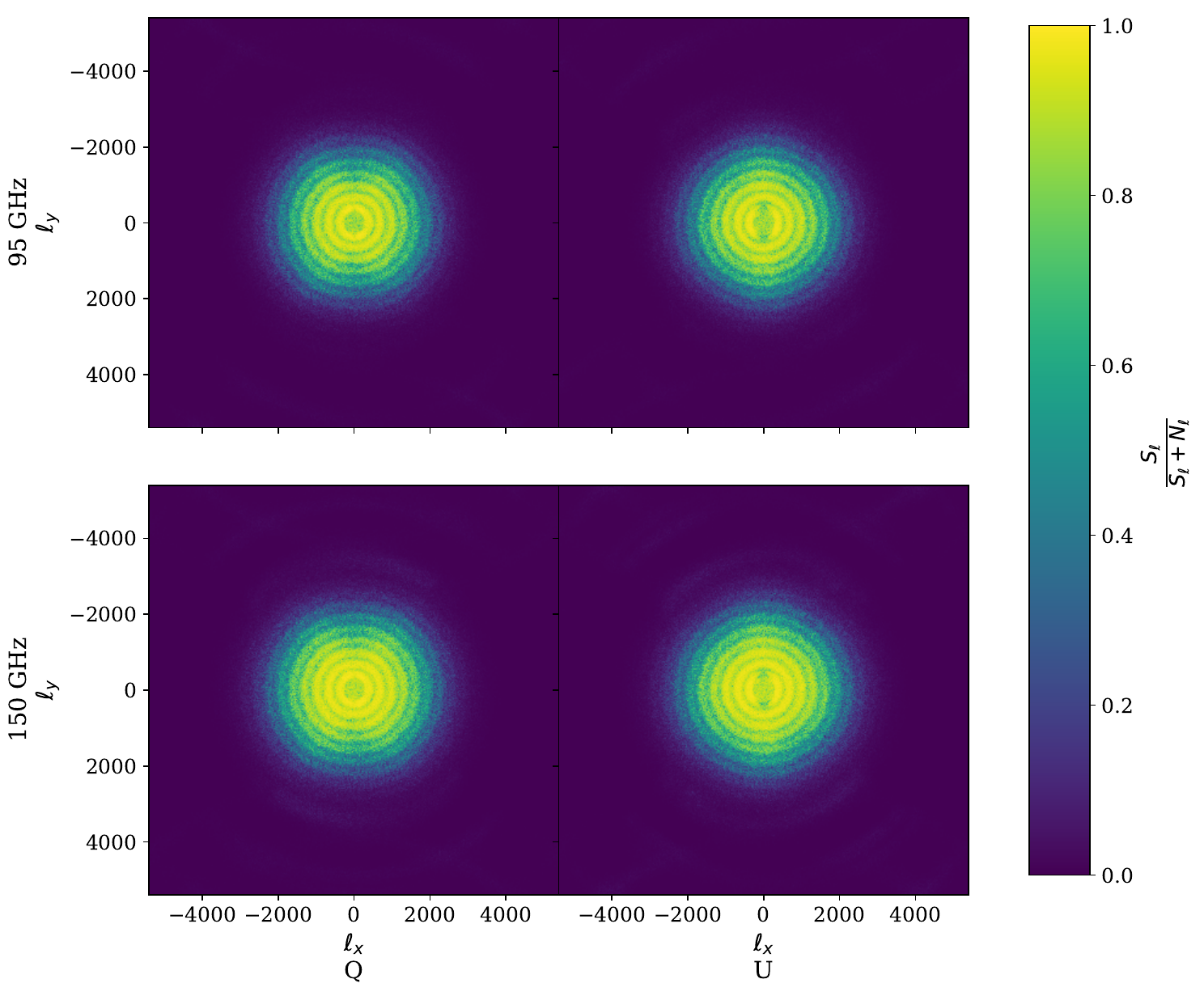}
\caption{$\ell$-space Wiener filter that is applied to the template coadds to downweight noisy modes. Because the noise properties of the maps differ between Stokes $Q$ (left column) and $U$ (right column), as well as between $95$ GHz (top row) and $150$ GHz (bottom row) observing bands, each must be filtered independently.}
\label{fig:wiener-filter}
\end{figure*}

\subsubsection{Template coadds}
As mentioned above, we use the full-season coadded $Q$ and $U$ maps as estimates for the true CMB polarization fields $Q_0$ and $U_0$. Although the maps are signal-dominated on most relevant scales, the noise contribution is not negligible; this noise biases the estimator for the angle $\hat{\rho}$. Given the noise level of our dataset, we observe a ${\sim}50$-$60\%$ reduction in the value of $\hat{\rho}$. To see why this bias occurs, consider the limit where $Q_0$ and $U_0$ are composed of only noise and no CMB. Due to the small-angle approximation made in our model, any rotation adds noise power in this limit and makes the $\chi^2$ larger, so is disfavored by the angle estimator.

To mitigate this bias, we apply a Wiener filter to the full-season coadds,

\begin{align}
\begin{split}
Q_0^\prime &= \mathcal{F}^{-1} \left\lbrace \mathcal{F}\left\lbrace Q_0 \right\rbrace \frac{S_{\ell,Q}}{S_{\ell,Q} + N_{\ell,Q}} \right\rbrace, \\
U_0^\prime &= \mathcal{F}^{-1} \left\lbrace \mathcal{F}\left\lbrace U_0 \right\rbrace \frac{S_{\ell,U}}{S_{\ell,U} + N_{\ell,U}} \right\rbrace,
\end{split}
\end{align}
where the prime denotes the filtered map, $\mathcal{F}$ denotes the Fourier transform, and $S_{\ell,P}$ and $N_{\ell,P}$ represent the two-dimensional signal (i.e., CMB) and noise power spectra, respectively, for Stokes parameter $P$. Because the noise properties vary by Stokes parameter and by observing band, each band's $Q$ and $U$ maps are filtered independently. These filters are shown in Fig. \ref{fig:wiener-filter}. They effectively down-weight noisy modes by emphasizing modes with high signal-to-noise ratios (that is, the same modes where the CMB EE power spectrum peaks).

$S_\ell$ was determined using the SPT-3G map-space simulation pipeline, which is described in brief here (see D21 for full details). In a process called \textit{mock observation}, fake TOD are generated from a simulated sky using the actual pointing, detector selection, and TOD weights from an observation. These TOD are passed through the entire mapmaking pipeline to create a simulated map.

To determine $S_\ell$, we created $10$ noise-free, Gaussian realizations of the CMB sky,  with underlying power spectra determined using the best-fit cosmological parameters from the \textsc{base\_plikhm\_ttteee\_lowl\_lowe\_lensing} $2018$ \textit{Planck} data release \citep{planck18-6}. Each realization was mock-observed with the pointing/detector-cutting information from three random observations per subfield, and the resulting $12$ maps were coadded together and the power spectrum was estimated. These $10$ power spectra were then averaged to give us $S_\ell$.

$N_\ell$ was estimated from the data themselves using season-long \textit{signflip noise realizations}. For every observation, we subtracted the left-going map from the right-going one to remove the CMB signal. The resulting difference map is then assigned a random sign and the full set is coadded together to give an estimate of the coadded noise for the full season. We generated $33$ of these noise realizations per observing band, took the power spectrum of each, and set $N_\ell$ equal to the average spectrum.

After filtering, $Q_0$ and $U_0$ are not perfect representations of the CMB, and this will leave some residual bias. We can use the same simulation framework to test whether this bias is at an acceptable level. We filtered a noisy template coadd and used it to estimate angles on a collection of noisy single-observation simulated maps with a $2.00$ degree $Q$/$U$ rotation injected. The distributions of reconstructed angles have mean $2.03 \pm 0.06$ ($2.02 \pm 0.05$) degrees at $95$ ($150$) GHz; thus, we conclude that using a Wiener-filtered template coadd reduces the bias caused by a noisy template coadd to a negligible level. However, the filtering comes at the cost of a sensitivity reduction of approximately $10\%$ (as measured by the magnitude of the uncertainty on $\hat{\rho}$).

There is another bias introduced by the use of the full-season coadds as the estimates for $Q_0$ and $U_0$. In the presence of a signal, the true $Q$ and $U$ fields are slightly washed out in the coadd, making the polarization rotation angle measured in individual maps appear larger than it truly is. However, this is a second-order effect (that is, it scales as ${O}\left({\phi_0}^2\right)$) and can be safely neglected here.\footnote{Washout during last scattering, as described in Sec. \ref{sec:intro}, is non-negligible because the strength of the axion field $\phi_0$ is much larger during last scattering than today.}

\subsubsection{Mapmaking procedure bias}
\label{sec:TF}
It is well documented that the TOD filtering biases the estimation of CMB power spectra \citep{hivon02}, a bias which must be accounted and corrected for in power spectrum analyses by determining the transfer function of the mapmaking procedure. This power spectrum bias does not bias the estimation of $\hat{\rho}$; it only adds a small amount of variance due to the removal of E-modes. However, it is possible that our mapmaking procedure could introduce a bias to $\hat{\rho}$ that should be corrected.

In order to test this, we again generated a set of noise-free Gaussian CMB realizations, applying a $Q$/$U$ rotation to these mock skies (arbitrarily chosen to be $2.0$ degrees) before mock-observing them with a random subset of observations. We observed a slight reduction in the value of the angle we reconstructed from these maps, on the order of $2\%$. It is unclear what the source of this bias is, but the F19 upper limits on $g_{\phi\gamma}$ place the amplitude of rotation to be $< 0.1$ deg; at this level the bias should be $< 0.002$ deg. Because this bias is entirely negligible when compared with the uncertainty on the angles from each observation (discussed in more detail in Sec. \ref{sec:uncertainty}), we elect to not correct for it. This is, however, a potential improvement to be made in future analyses of this type.

\subsubsection{Map-space source masking}
\label{sec:masking}
As described in Sec. \ref{sec:TOD-maps} and in more detail in D21, timestream samples where a bolometer is pointed at a point source are masked during TOD filtering. This avoids the creation of artifacts from the polynominal filtering of the timestream but leaves the sources themselves in the output maps. The sources' time-varying polarization power can bias the estimated angles in a way that looks like a false axion signal and causes jackknife failures. For example, PMN 0208-512 is a bright, variable AGN in the SPT-3G survey area whose flux varies between ${\sim}1$-$5$ Jy and produces a detectable time variation in our polarization angle estimator. To account for the bias from sources like this, we apply a map-space mask with a $5$-arcminute radius to all sources detected above $50$ mJy in a coadd of $95$ GHz data from SPT-3G's $2018$ observing season (though the list is chosen based on source flux at $95$ GHz, the same sources are masked when calculating angles for both observing bands). Once the mask is applied, we calculate $\hat{\rho}$ for each observation. This threshold was chosen based on \citet{henning18}, which demonstrated that sources below the cutoff flux value contribute negligible power to polarization power spectra. We confirmed that the variance added to $\hat{\rho}$ by leaving these dim sources unmasked is subdominant to the intrinsic uncertainty in the estimate (Sec. \ref{sec:uncertainty}).

We end up masking ${\sim} 2\%$ of the effective sky area in the SPT-3G field. The uncertainty on the final rotation amplitude scales approximately as the inverse-square-root of the sky fraction observed (Sec. \ref{sec:results}), so this masking leads to a sensitivity loss of only ${\sim} 1\%$.

\subsubsection{Estimating the uncertainty on $\hat{\rho}$}
\label{sec:uncertainty}
To estimate the uncertainty on the polarization rotation angle for each observation, we require a method to generate many noise realizations with the statistical properties of the noise in that particular observation's map. We calculate an angle for each of these noise realizations, and set the uncertainty on $\hat{\rho}$ to be $\sigma_{\hat{\rho}}$, the standard deviation of the distribution of angles.

We take inspiration from the season-long signflip noise realizations detailed in Sec. \ref{sec:coadds} and devise a method of generating signflip noise realizations on the per-observation level. For each scan pair, we subtract one scan from the other, leaving only a noise estimate for that scan pair. We then assign a random sign to each pair's noise map and coadd all $36$ scan pairs together to get a noise realization for the full observation. We generate $1000$ such realizations per observation, allowing us to determine the uncertainty with high precision. The average uncertainty on $\hat{\rho}$ is $2.50$ deg for $95$ GHz observations and $2.01$ deg for $150$ GHz observations.

\begin{figure*}
\includegraphics[width=2\columnwidth]{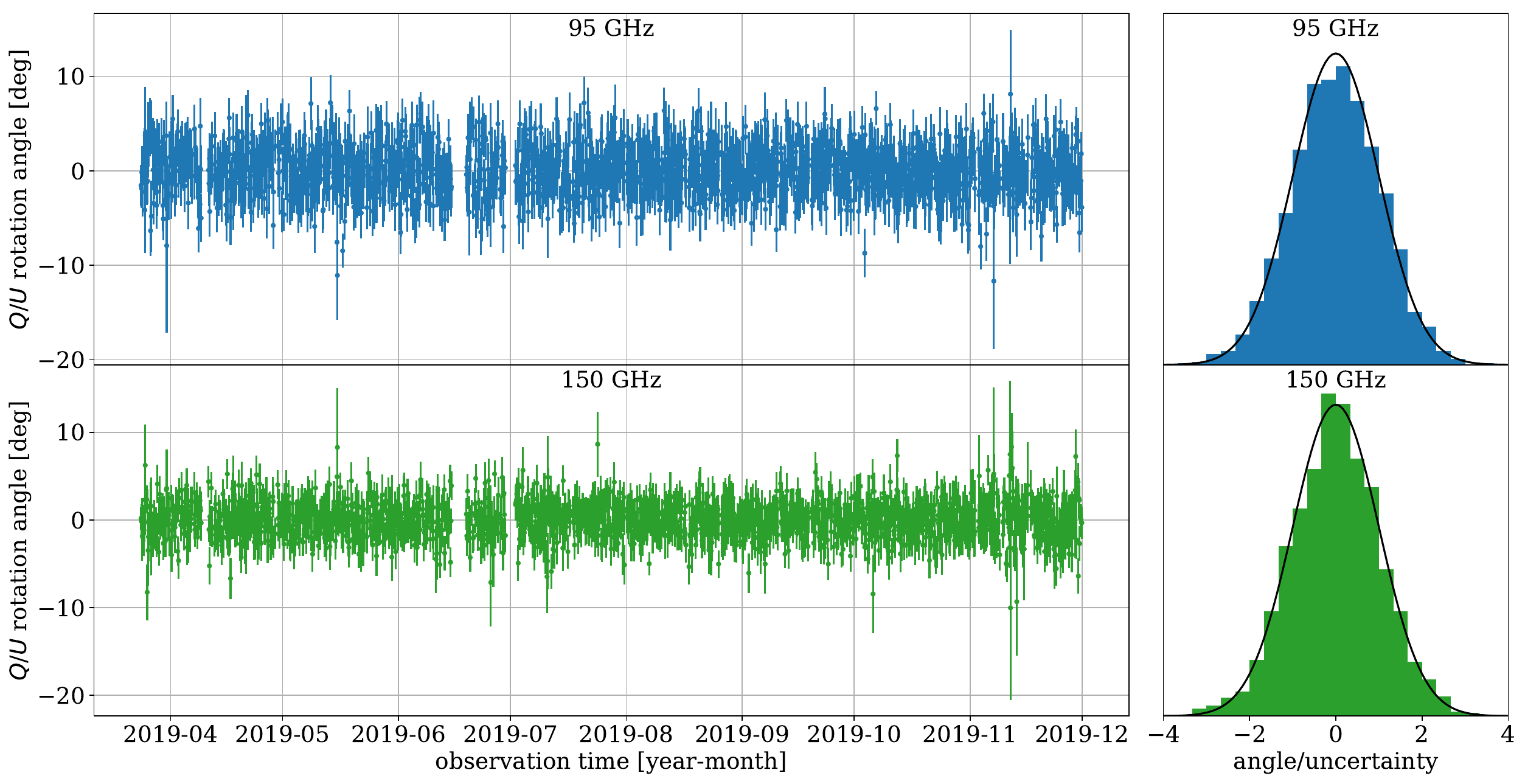}
\caption{(left) Time series of polarization rotation angles measured for both the $95$ GHz and $150$ GHz bands. The gaps where there are no angles for a short period correspond with telescope downtime due to unscheduled drive maintenance. (right) Histograms of $x = \hat{\rho} / \sigma_{\hat{\rho}}$ for both observing bands. In both cases, this quantity is consistent with a unit Gaussian, plotted as a solid black line.}
\label{fig:timeseries}
\end{figure*}

With our chosen TOD filtering settings, we expect our single-observation maps to be dominated by white noise. Therefore we also expect the quantity $x = \hat{\rho} / \sigma_{\hat{\rho}}$ to be Gaussian distributed with mean zero and standard deviation unity. As a consistency check that we are estimating $\sigma_{\hat{\rho}}$ correctly, we perform a Kolmogorov-Smirnov (KS) test for Gaussianity on these distributions for both bands. We find a p-value on the KS test of $0.621$ ($0.877$) for the $95$ ($150$) GHz data. Because these are within the $95$th percentile, we claim that $x$ is consistent with being Gaussian-distributed. The final time series of angles and uncertainties are shown for both observing bands in Fig. \ref{fig:timeseries}.

\subsection{Jackknives}
\label{sec:jackknives}
Once we have a time series of polarization rotation angles, we perform a suite of jackknife tests to search for systematic effects in the data. These tests can be broken up into three categories: \textit{temporal jackknives}, for binary quantities that vary in time (such as whether the moon is above or below the horizon); \textit{continuous jackknives}, for continuous quantities that vary in time (such as observation azimuth); and \textit{null jackknives}, for data combinations where we expect the signal to be nulled (such as left-right difference maps).

All of the jackknife tests depend in some way on simulated time series of polarization rotation angles. For each observation $i$ in the fake time series, we simulate an angle $\rho_{\textrm{sim},i}$ by randomly selecting an observation $j$ from the same subfield as observation $i$ and computing

\begin{equation}
\rho_{\textrm{sim},i} = \hat{\rho}_j \frac{\sigma_{\hat{\rho},i}}{\sigma_{\hat{\rho},j}}.
\end{equation}
Each quantity on the right-hand side of the equation comes from the actual data; in this way we are able to create simulated time series with noise properties consistent with those of the real data.

\subsubsection{Temporal jackknives}
\label{sec:temp-jack}
We use the temporal jackknife to test for systematics induced by quantities that take on one of two distinct values in each observation. Specifically, we split our time series in three ways:
\begin{itemize}
\item[(i)] Sun up/down, to test for sun contamination in the data through telescope sidelobes.
\item[(ii)] Moon up/down, to test for false signals from the periodic rise and fall of the moon.
\item[(iii)] An elevation-based test that compares data from two different subfields, for all possible subfield pairs, in order to probe atmospheric effects.
\end{itemize}

In each case, we construct a likelihood

\begin{equation}
\label{eqn:likelihood}
\mathcal{L}\left( A, \delta, f \right) = \exp\left( -\sum_i \frac{\left( \hat{\rho}_{i} - \rho_{\textrm{m},i} \right)^2}{2 \sigma_{\hat{\rho},i}^2} \right) \equiv \exp\left( -\frac{\chi^2_\textrm{TS}}{2} \right),
\end{equation}
where $\rho_{\textrm{m},i} = A \sin\left( 2\pi f t_i + \delta \right)$ is the model angle [Eq. (\ref{eqn:angle-model})] at time $t_i$ and the summation is over observations. $\chi^2_\textrm{TS}$ is the time series $\chi^2$, not to be confused with the map-space $\chi^2$ introduced in Eq. (\ref{eqn:chi2}). Then we take as a test statistic $\lambda_t$, defined to be the log-likelihood ratio

\begin{widetext}
\begin{equation}
\lambda_t \equiv -2\log \left( \frac{\text{max}_{A,\delta, f} \left[ \mathcal{L}_{tot} \left( A, \delta, f \right)  \right]}{\left[\text{max}_{A_1,\delta_1} \mathcal{L}_1 \left( A_1, \delta_1 \right) \right] \left[\text{max}_{A_2,\delta_2} \mathcal{L}_2 \left( A_2, \delta_2 \right) \right]} \right),
\end{equation}
\end{widetext}
where $\mathcal{L}_{tot}$ is the likelihood of the full time series and $\mathcal{L}_i$ is the likelihood of the $i$th split time series. In the $\mathcal{L}_i$ functions, the frequency has been fixed to the best-fit frequency from the full likelihood optimization, as this caused the distribution of $\lambda_t$ values to be closer to a $\chi^2$ distribution. This frequency-fixing is a valid nested hypothesis, such that the likelihood ratio continues to be an optimal test statistic, albeit over a reduced parameter space. With this definition, $\lambda_t$ will be large in cases where there is an oscillatory systematic in one of the two splits. We consider frequencies between $0.01$ inverse-days and $2.00$ inverse-days, with a frequency spacing of $5 \times 10^{-4}$ inverse-days. This frequency spacing oversamples the frequency width of a sine wave, ensuring that we are sensitive to all possible signals in the considered range. The test statistic for the data is compared to a distribution of test statistics from simulated background-only time series in order to calculate a p-value.

Due to the frequency fixing, the temporal jackknife is only sensitive to systematics at the best-fit frequency for the full time series. We are especially interested in testing for systematics at this frequency because this is where a potential signal is likely to appear. However, due to windowing effects (Sec. \ref{sec:angs-lims}), and because we wish to set limits at all frequencies under consideration, we search for systematic effects at other frequencies as well. In order to do so, we also perform a variation on the temporal jackknife that we denote the \textit{noise jackknife}. In the noise jackknife tests, the best-fit signal is subtracted from the full time series. Then the slightly altered log-likelihood ratio

\begin{widetext}
\begin{equation}
\lambda_{n,i} = -2\log \left( \frac{\text{max}_{A, \delta} \left[ \mathcal{L}_{tot} \left( A, \delta, f_i \right) \right]}{\left[\text{max}_{A_1, \delta_1} \mathcal{L}_1 \left( A_1, \delta_1, f_i \right) \right] \left[\text{max}_{A_2, \delta_2} \mathcal{L}_2 \left( A_2, \delta_2, f_i \right) \right]} \right)
\end{equation}
\end{widetext}
is computed at all $3981$ frequencies $f_i$ under consideration. To pare this information down to a single p-value, we compute the test statistic $\lambda_n$, defined as
\begin{equation}
\lambda_n \equiv \text{max}_i \left( \lambda_{n,i} \right),
\end{equation}
and compare this with a distribution of similar test statistics from simulated background-only time series.

\subsubsection{Continuous jackknives}
\label{sec:cont-jack}
SPT-3G's location at the South Pole, coupled with the fact that it observes a patch of fixed RA in the sky, means that observations are taken across the entire $2\pi$ range in azimuth. If there is a systematic induced by \textit{ground pickup} (that is, light scattering off of ground-based features), it ought to show up as a function of azimuth. Though this is a temporally varying quantity, we cannot use the temporal jackknife since azimuth takes on continuous rather than binary values. Thus we implement the continuous jackknife to test for azimuth-synchronous signals.

Before running this test, the best-fit signal in time is subtracted from the time series. We then fit a sinusoid to the time series as a function of observation azimuth rather than time. Its amplitude is compared to a distribution of amplitudes from simulated background-only time series in order to calculate a p-value. We choose to look only at the fundamental mode (that is, an azimuthal sinusoid with a period of $2\pi$) and to neglect higher-frequency azimuthal modes because the horizon around the SPT is mostly featureless, with the exception of the Dark Sector Laboratory building where the SPT is housed. Although this feature will not appear as a pure sine wave, the strongest component of its Fourier decomposition will be the fundamental mode and thus this test is sensitive to the most likely cause of ground pickup.

\subsubsection{Null jackknives}
\label{sec:null-jack}
\begin{center}
\begin{table*}
\def\arraystretch{1.5}
\setlength{\tabcolsep}{10pt}
\begin{tabular}{ c|c c|c c }
\hline\hline
 & \multicolumn{2}{c|}{95 GHz} & \multicolumn{2}{c}{150 GHz} \\
 \hline
 & Temporal & Noise & Temporal & Noise \\
 \hline
 Moon up/down & 0.1865 & 0.5159 & 0.9248 & 0.7545 \\
 Sun up/down & 0.4366 & 0.6819 & 0.4146 & 0.7681 \\
 el0/el1 & 0.3338 & 0.1424 & 0.8984 & 0.6317 \\
 el0/el2 & 0.2566 & 0.7275 & 0.0067 & 0.5979 \\
 el0/el3 & 0.0854 & 0.1047 & 0.0123 & 0.0808 \\
 el1/el2 & 0.9482 & 0.0746 & 0.7213 & 0.3605 \\
 el1/el3 & 0.4019 & 0.3103 & 0.7865 & 0.6122 \\
 el2/el3 & 0.0828 & 0.4516 & 0.7932 & 0.4133 \\
 \hline
 Azimuthal & \multicolumn{2}{c|}{0.6066} & \multicolumn{2}{c}{0.0271} \\
 \hline
 Null & \multicolumn{2}{c|}{0.0655} & \multicolumn{2}{c}{0.8561} \\
 \hline
 95 GHz / 150 GHz & \multicolumn{4}{c}{0.9992} \\
 \hline
\end{tabular}
\caption{P-values for the full suite of jackknife tests performed to search for evidence of systematics in the time series of polarization rotation angles. The minimum p-value of $0.0067$ is greater than our success threshold of $0.05 / N_\mathrm{tests} = 0.0014$, and the p-value on a KS test for uniformity is $0.4416$, greater than our success threshold of $0.05$. While the p-value for the $95$ GHz / $150$ GHz jackknife test is unusually high, this signifies that the data are even more consistent with displaying no systematic signal than expected. Therefore we find no evidence of significant systematic effects.}
\label{tab:pvalues}
\end{table*}
\end{center}

This final jackknife test was developed to search for systematic signals in quantities where any true axionlike signal should be nulled. It is used to probe scan-direction-dependent systematic effects (as could be caused by our decision to not correct for detector time constants) as well as differences between the $95$ GHz and $150$ GHz observing bands (as could be caused by astrophysical foregrounds). We do not expect any systematics in these quantities, so these tests serve as an internal consistency check.

First, a time series is constructed of angles with the expected signal nulled. In the scan-direction case, this involves calculating a polarization angle from maps where left-going and right-going scans have been given opposite signs. In the observing band case, it involves subtracting the two time series (while adding their uncertainties in quadrature). Once we have the null time series, we compute the amplitude of the best-fit sinusoid at every frequency. Similarly to the noise jackknife, we take as a test statistic the largest of these amplitudes. A p-value is then computed by comparing with a distribution of test statistics from simulated background-only time series.

\subsubsection{Jackknife results}
\label{sec:jack-results}
We set two criteria to determine whether we pass our jackknife tests. First, the smallest p-value must be larger than $0.05 / N_\mathrm{tests}$, or $0.0014$ with our $37$ tests. Second, we expect the distribution of all p-values to be uniform in the absence of systematics, so we perform a KS test for uniformity and require that the p-value on this KS test be greater than $0.05$.

The full suite of p-values is presented in Table \ref{tab:pvalues}. The smallest p-value is $0.0067$ and the p-value on the KS test for uniformity is $0.4416$. Thus we pass our jackknife tests and conclude that there is no evidence of strong systematic effects in the data.

\subsection{Angles to upper limits}
\label{sec:angs-lims}
Once we have a time series of polarization rotation angles, the next step is to calculate upper limits on the polarization rotation amplitude. This is done independently for every frequency/mass bin. As stated before, we consider frequencies spaced $5 \times 10^{-4}$ inverse-days apart between $0.01$ inverse-days and $2.00$ inverse-days (or, in terms of oscillation period, between $12$ hours and $100$ days). Our data points are unevenly spaced roughly $2$ hours apart and span a range of just over $250$ days, allowing us to sample the full oscillation over the course of the season (the consequences of this uneven sampling are discussed in Sec. \ref{sec:window}).

\begin{figure*}
\includegraphics[width=2\columnwidth]{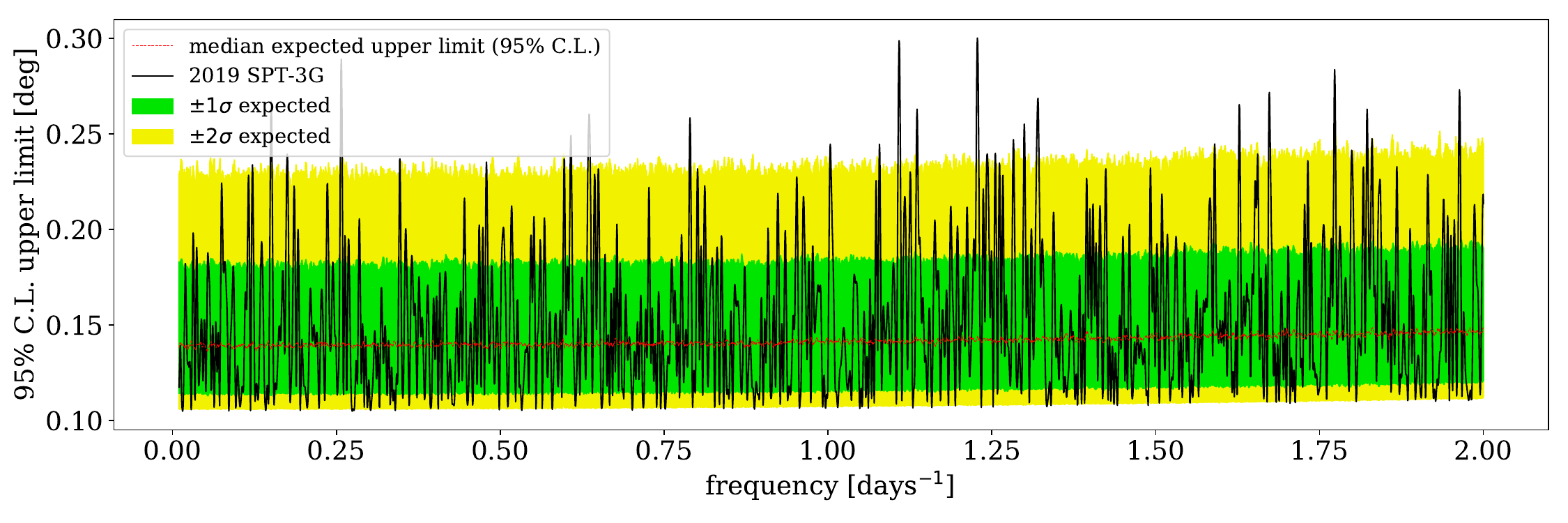}
\caption{$95 \%$ C.L. upper limits on $Q$/$U$ rotation angle as a function of oscillation frequency (solid black line), along with simulated background-only median behavior (red) and $1\sigma$ (green) and $2\sigma$ (yellow) regions. As described in Sec. \ref{sec:maps-angs}, averaging over the course of an observation leads to less stringent limits as the oscillation frequency increases. However, this is a small effect; it is on the order of only ${\sim} 5\%$ at $2.00$ inverse-days. Due to the large number of frequency bins, we expect some limits in excess of the $2\sigma$ background contour; this does not necessarily constitute evidence for a sinusoidal polarization rotation.}
\label{fig:upper-lims}
\end{figure*}

To set an upper limit at a fixed frequency $f_0$, we first construct a likelihood like the one defined in Eq. (\ref{eqn:likelihood}), except that the sum is over all observations and observing bands. That likelihood is marginalized over the phase $\delta$,
\begin{equation}
\mathcal{L}_\textrm{m}\left( A \right) = \int_{0}^{2\pi} \mathcal{L}\left( A, \delta, f_0 \right) d\delta.
\end{equation}
We assume a uniform prior on amplitude with $A_\textrm{max} = 0.5$ deg,\footnote{As long as the upper bound on the prior is high enough, the result is insensitive to the exact choice since the weight is concentrated at low amplitude.}
\begin{align}
\tilde{P}\left( A \right) = \begin{cases}
							\frac{1}{A_\textrm{max}}, \quad &0 < A < A_\textrm{max}, \\
							0, &\mathrm{otherwise},
							\end{cases}
\end{align}
and use this prior to construct a posterior probability distribution,
\begin{equation}
P\left( A \right) = \frac{\tilde{P}\left( A \right) \mathcal{L}_\textrm{m}\left( A \right)}{\int_{0}^{A_\textrm{max}} \tilde{P}\left( A^{\prime} \right) \mathcal{L}_\textrm{m}\left( A^{\prime} \right)dA^{\prime}}.
\end{equation}
This is integrated to obtain a cumulative density function,
\begin{equation}
F\left( A \right) = \int_{0}^{A} P\left( A^{\prime} \right) dA^{\prime}.
\end{equation}
The upper limit $\tilde{A}$ at a given confidence level is then the amplitude at which the CDF is equal to said confidence level (taken to be $0.95$ here). The upper limits set by this analysis, as well as the background-only model contours, are shown in Fig. \ref{fig:upper-lims}. The median expected limit is nearly constant as a function of frequency, but degrades slightly at higher frequencies due to a changing rotation angle over the course of the ${\sim} 2$-hour observation [Eq. (\ref{eqn:average})]. As described in Sec. \ref{sec:maps-angs}, the limit would also degrade for low frequencies, though we do not consider any frequencies low enough for this to take effect. Below $1.00$ inverse-days, where the effect of averaging is negligible (that is, $\lesssim 1\%$), we set a median limit of

\begin{equation}
\label{eqn:qu-lim}
\tilde{A} < 0.142 \text{ deg},
\end{equation}
corresponding to $A_\textrm{sky} < 0.071$ deg.

\subsubsection{Observation window function}
\label{sec:window}
\begin{figure}
\includegraphics[width=\columnwidth]{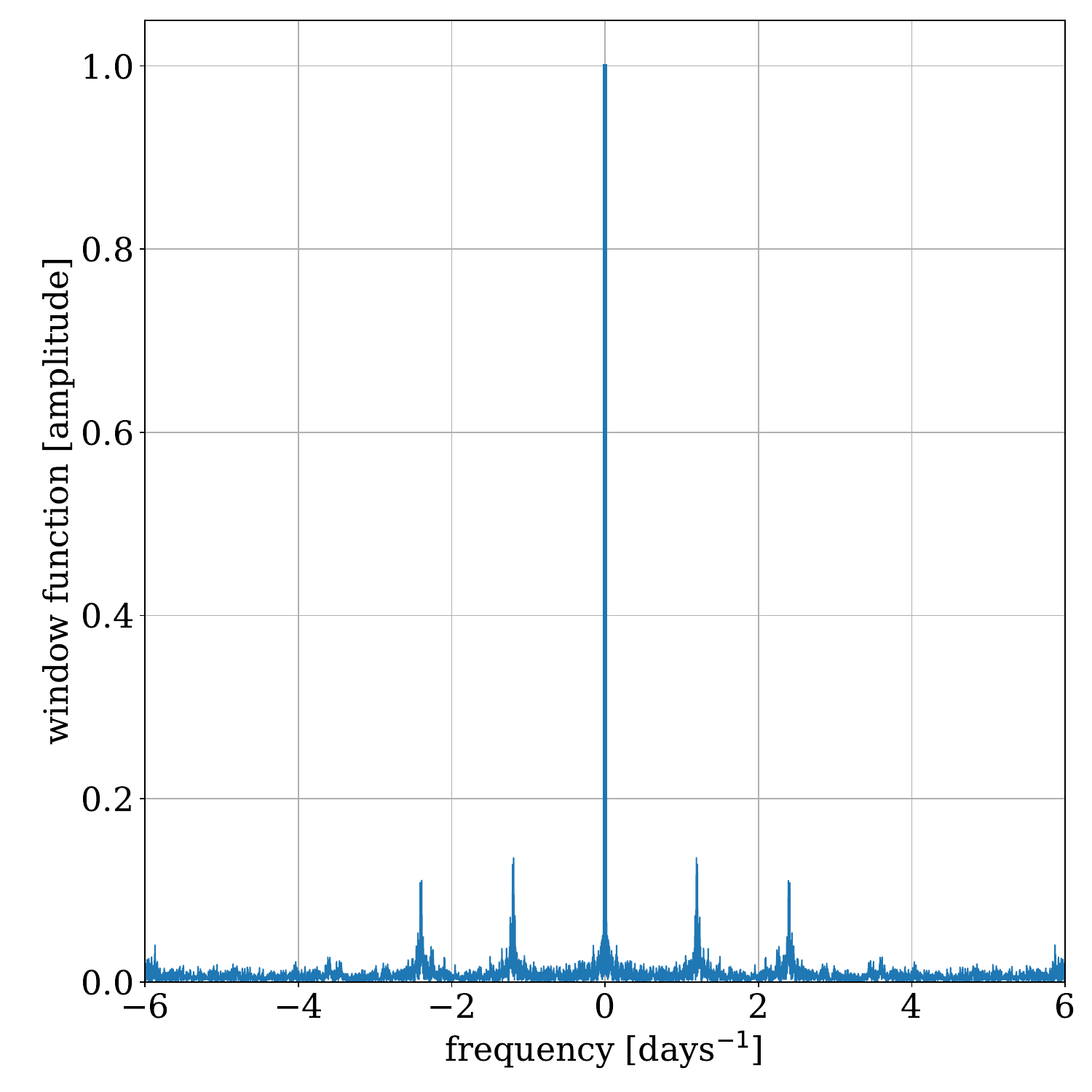}
\caption{Window function (in amplitude) of the observation times used in this analysis, which characterizes the extent to which signals at a single frequency produce detectable power at other frequencies in our likelihood analysis.
The largest sidelobes are at an amplitude of $14\%$ of the main lobe, and these result from the quasiperiodic pattern of sorption refrigerator cycles and observations that occur between them in time.
In an analysis such as ours without a large expected signal, the presence of sidelobes at this level does not impact the interpretation of our results.}
\label{fig:window_function}
\end{figure}

During the course of the season, observations do not occur at equally spaced intervals in time, so the times that we assign to polarization angles in the likelihood of Eq. (\ref{eqn:likelihood}) are not uniformly spaced.
Although observations occur on a scheduled cadence between recharging the sorption cooler every ${\sim} 19.5$ hours, the schedule within this period combines CMB subfield observations with different types of calibration observations, and the frequency with which each subfield was observed was furthermore adjusted throughout the season.
In our likelihood analysis, the irregular sampling behaves similarly to a window function that is convolved with sinusoidal signals in the data.
Since the sampling is not uniform, the window function can in principle have power at any frequency, unlike the Dirac comb window function that corresponds to uniform sampling.
When convolved with a sinusoid at a fixed frequency $f$, this may cause us to detect signals at frequencies other than $f$.
This behavior is well documented in similar methods that identify sinusoidal signals in irregularly sampled data, such as the Lomb-Scargle periodogram \citep{vanderplas18}.

While this windowing phenomenon does affect our analysis, it can be practically neglected because of the structure of the SPT-3G window function.
The window function (in amplitude) of the observation times is given by
\begin{equation}
W (f) = \left| \sum_{j=0}^{N} \exp \left( -2\pi i f t_j \right) \right|,
\end{equation}
where the $t_j$ are the times of the $N$ observations in our dataset and $f$ is frequency.
Figure \ref{fig:window_function} shows the window function for our data. The majority of power is in the central lobe and two symmetric sidelobes at the level of $14\%$ of the main lobe in amplitude.
The analysis would therefore have to detect a signal at high significance before sidelobes were to be detectable, and these sidelobes furthermore would occur at predictable frequency offsets from the main signal.
Given the existing constraints from the \textit{Planck} washout analysis \citep{fedderke19}, we do not expect to detect a signal with high significance in the present work, and any sidelobes due to the window function will be subdominant to noise.

One further possible impact of the window function structure is that systematics that induce oscillation of the polarization angle at frequencies \emph{outside} our search band of $0.01$ inverse-days to $2.00$ inverse-days could have sidelobes that appear as signals \emph{inside} our search band.
The jackknife tests described in Sec. \ref{sec:jackknives}, however, are sensitive to these in-band sidelobes from out-of-band systematic effects, so the impact of this phenomenon is only to complicate the physical interpretation of failures of the jackknife tests.


\section{Results and Discussion}
\label{sec:results}
Although the $95 \%$ C.L. data limit in Fig. \ref{fig:upper-lims} exceeds the $2\sigma$ background contour in a number of frequency bins, this is not necessarily evidence of a time-varying birefringence signal due to the large number of frequency bins under consideration. We test for detection of such a signal in a similar manner to BK22. For each frequency $f_i$, we compute the quantity
\begin{equation}
\Delta \chi^2_{\textrm{TS},i} = \chi^2_\textrm{TS} \left( 0,0 \right) - \chi^2_{\textrm{TS},i} \left( A_0, \delta_0 \right),
\end{equation}
where the subscript $0$ signifies the value that minimizes the $\chi^2$ for that frequency bin. We take as a test statistic 
\begin{equation}
\lambda_{\chi^2_\textrm{TS}} \equiv \text{max}_i \left(\Delta \chi^2_{\textrm{TS},i} \right).
\end{equation}
A p-value testing for consistency with background is then determined by comparing $\lambda_{\chi^2}$ from data with a distribution of similar test statistics computed from background-only simulations. Using this method, we find that the data are consistent with the background-only model with a p-value of $0.48$.

\begin{figure*}
\includegraphics[width=2\columnwidth]{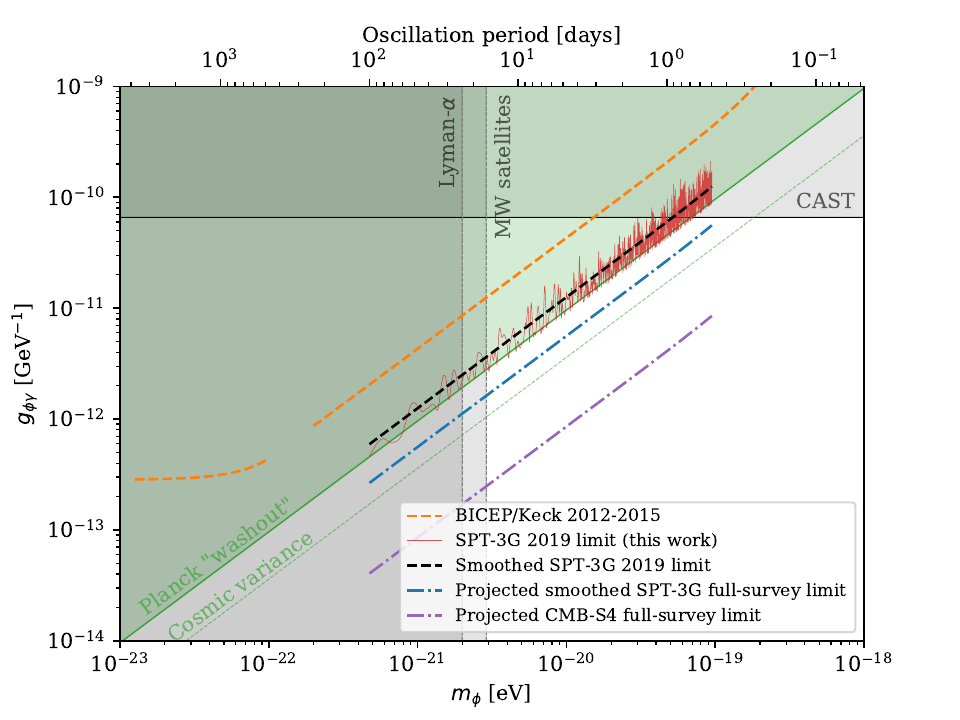}
\caption{The parameter space for axion-photon coupling $g_{\phi\gamma}$ as a function of axion mass $m_{\phi}$. The SPT-3G $95\%$ C.L. upper limit is given by the solid red line, and the smoothed fit to this [Eq. (\ref{eqn:final-lim})] by the dashed black line. The dashed orange lines represent the most recent limits set by the BICEP/Keck collaboration \citep{bicepkeck22} using the AC oscillation effect; the solid green line represents the limit set with \textit{Planck} data using the washout effect, with the dashed green line providing the strongest possible limit that could be set with the washout effect due to cosmic variance \citep{fedderke19}. Projected limits using the AC oscillation effect for the full SPT-3G survey as well as the future CMB-S4 survey are given by the dot-dashed blue and purple lines, respectively (although these projections do not account for the wider mass range that full survey analyses could constrain). The CAST limit on $g_{\phi\gamma}$ \citep{cast17} is given by the horizontal gray line (stronger limits set using data from the supernova SN1987A \citep{payez15}, and \textit{Chandra} X-ray spectroscopy \citep{reynolds20} are excluded from the plot as a result of difficult-to-quantify modeling uncertainties). Lower mass limits from observations of Lyman-$\alpha$ emission \citep{irsic17} and Dark Energy Survey observations of Milky Way satellite galaxies \citep{nadler21} are given by the labeled vertical dashed gray lines (while stronger limits from Lyman-$\alpha$ observations exist \citep{rogers21}, we have chosen to plot a more conservative limit). Both the BICEP/Keck and the SPT-3G results assume that axions comprise the entirety of the local dark matter, and that the density of the local dark matter field is $0.3$ GeV/cm$^3$.}
\label{fig:param-limit}
\end{figure*}

The upper limit on rotation amplitude can be converted into an upper limit on the axion-photon coupling constant $g_{\phi\gamma}$ following the method in F19:

\begin{align}
\begin{split}
g_{\phi\gamma,\mathbf{UL}} &= \left( 2.1 \times 10^{9} \text{ GeV} \right)^{-1} \times \tilde{A} \\
         &\times \left( \frac{m_{\phi}}{1.0 \times 10^{-21} \text{ eV}} \right) \times \left( \kappa \frac{\rho_0}{0.3 \text{ GeV/cm}^3} \right)^{-1/2},
\end{split}
\end{align}
where $\tilde{A}$ is the measured upper limit on $Q/U$ rotation amplitude in radians, $\kappa$ is the fraction of local dark matter comprising axions, and $\rho_0$ is the density of the local dark matter field. Recalling the degradation in sensitivity at higher frequencies due to the noninstantaneous sampling of the polarization rotation angle [Eq. (\ref{eqn:average})], we can fit a smoothed approximation to these limits of the form

\begin{equation}
A < \frac{A_0}{\sinc\left( m_{\phi}\Delta t / 2 \right)}
\end{equation}
with $A_0$ as a free parameter and $\Delta t=2.13$ hours the mean observation duration. Performing a least-squares fit to the determined limits $\tilde{A}$, we find $A_0 = 0.151$ deg. If we assume that the local dark matter density is $0.3$ GeV/cm$^3$ and that axions comprise the full fraction of the dark matter, this translates to
\begin{align}
\label{eqn:final-lim}
\begin{split}
g_{\phi\gamma} < 1.25 &\times 10^{-12}\text{ GeV}^{-1} \times \left( \frac{m_{\phi}}{1.0 \times 10^{-21} \text{ eV}} \right) \\
                      &\times \left[ \sinc\left( \frac{m_{\phi}}{1.72 \times 10^{-19} \text{ eV}} \right) \right]^{-1}.
\end{split}
\end{align}

This limit on $g_{\phi\gamma}$ is shown for our results, along with other relevant limits in this region of parameter space, in Fig. \ref{fig:param-limit}. For frequencies below $1.00$ inverse-days, where the limit is approximately flat, we take the approximation in Eq. (\ref{eqn:qu-lim}) to set a median limit of

\begin{equation}
g_{\phi\gamma} < 1.18 \times 10^{-12}\text{ GeV}^{-1} \times \left( \frac{m_{\phi}}{1.0 \times 10^{-21} \text{ eV}} \right).
\end{equation}
With a single year of data, SPT-3G sets the strongest limit yet using the AC oscillation effect, approximately $3.8$ ($3.4$) times stronger than BK22 for the flat (complete) region. At some masses this work sets the strongest limit of any CMB analysis yet, surpassing the washout limit set with \textit{Planck} polarization power spectra \citep{fedderke19}.

As a consistency check, we model the expected sensitivity difference between BK22 and the current work. In a simplified model, we expect the uncertainty to scale as

\begin{equation}
\label{eqn:sensitivity-model}
\sigma_\rho \propto n \times f_\textrm{sky}^{-1/2} \times \left( \sum_\ell C_\ell B_\ell \left( 2\ell + 1 \right) \right)^{-1/2},
\end{equation}
where $n$ is the combined noise level for all bands in the coadded template map, $f_\textrm{sky}$ is the fraction of the sky observed, and the final term is a scaling factor related to the size of the beam (and therefore the number of polarization modes each experiment is sensitive to). For BK22, the sky area is $400$ deg$^2$ and $n$ (in temperature) is approximately $1.8$ $\mu$K-arcmin \citep{bicep2keck18}; for this work, the sky area is $1500$ deg$^2$ and estimates accurate at the ${\sim} 10 \%$ level place $n$ at $4.4$ $\mu$K-arcmin. Finally, the current work is sensitive to approximately $16$ times as many modes as BK22. Given this, our toy model predicts SPT-3G to set a limit $3.2$ times stronger than BK22. Given the differences in analysis methods between the two limits, the uncertainty in the SPT-3G noise level, and the fact that BK22 used a somewhat reduced set of data when compared with \citep{bicep2keck18}, we find the true relative sensitivity to be in good agreement with this simple estimate.

When comparing these limits with others in the same region of parameter space, it is important to keep in mind that the limits set by F19, BK22, and the current work assume that the local dark matter is composed entirely of a single species of axion. If instead there are multiple axions, or a single type of axion makes up only a fraction of the local abundance, the limits become less stringent. The CAST limit \citep{cast17} is set strictly by Primakoff conversion of solar axions and is thus independent of any properties of local dark matter. While stronger limits on $g_{\phi\gamma}$ have been set in this mass range by observations of the supernova SN1987A \citep{payez15} and \textit{Chandra} X-ray spectroscopy \citep{reynolds20}, these limits are subject to large uncertainties stemming from source luminosity and magnetic field modeling, and are thus excluded from the plot. Conversely, the mass limits set by small-scale structure \citep{hui17}, Lyman-$\alpha$ emission \citep{irsic17, rogers21}, and Milky Way satellite galaxies \citep{nadler21} are wholly independent of the axion details, and only assume that an ultralight particle is the principle dark matter component. If the axions comprise some subdominant fraction of the dark matter, they could take on masses below this limit.

We reiterate that the current work uses only a single year of SPT-3G data. Since the sensitivity scales roughly as the inverse-square-root of the number of observations, we expect that a future analysis of this type using the full $5$-year SPT-3G dataset will improve the limits by more than a factor of two (as well as extend to a lower frequency range due to the longer observing time). Looking further ahead, the CMB field will begin capturing data with next-generation experiments such as Simons Observatory and CMB-S4. These experiments are expected to be much more sensitive to AC birefringence-type effects; estimates of such future limit-setting abilities are shown with the dot-dashed lines in Fig. \ref{fig:param-limit}. Due to the cosmic variance limit on axion searches using the polarization washout effect, it is the AC oscillation effect that will provide the strongest constraining power from CMB data on this type of measurement. Given that this is a relatively open region of parameter space, this means that there is a significant discovery potential in the future.


\section*{Acknowledgments}
The South Pole Telescope program is supported by the National Science Foundation (NSF) through Grants No. PLR-1248097 and No. OPP-1852617.
Partial support is also provided by the NSF Physics Frontier Center Grant No. PHY-1125897 to the Kavli Institute of Cosmological Physics at the University of Chicago, the Kavli Foundation, and the Gordon and Betty Moore Foundation through Grant No. GBMF\#947 to the University of Chicago.
Argonne National Laboratory's work was supported by the U.S. Department of Energy, Office of High Energy Physics, under Contract No. DE-AC02-06CH11357.
Work at Fermi National Accelerator Laboratory, a DOE-OS, HEP User Facility managed by the Fermi Research Alliance, LLC, was supported under Contract No. DE-AC02-07CH11359.
The Cardiff authors acknowledge support from the UK Science and Technologies Facilities Council (STFC).
The IAP authors acknowledge support from the Centre National d'\'{E}tudes Spatiales (CNES).
M.A. and J.V. acknowledge support from the Center for AstroPhysical Surveys at the
National Center for Supercomputing Applications in Urbana, IL.
J.V. acknowledges support from the Sloan Foundation.
K.F. acknowledges support from the Department of Energy Office of Science Graduate Student Research (SCGSR) Program.
The Melbourne authors acknowledge support from the Australian Research Council's Discovery Project scheme (No. DP210102386). 
The McGill authors acknowledge funding from the Natural Sciences and Engineering Research Council of Canada, Canadian Institute for Advanced Research, and the Fonds de recherche du Qu\'ebec Nature et technologies.
The UCLA and MSU authors acknowledge support from NSF AST-1716965 and CSSI-1835865.
This research was done using resources provided by the Open Science Grid \citep{pordes07, sfiligoi09}, which is supported by the NSF Award No. 1148698, and the U.S. Department of Energy's Office of Science.
The data analysis pipeline also uses the scientific python stack \citep{hunter07, jones01, vanDerWalt11}.

\bibliography{spt}

\end{document}